\documentclass[a4paper, 12pt, oneside, openright]{article}
\usepackage[T1]{fontenc}
\usepackage[utf8]{inputenc}
\usepackage[english]{babel}
\usepackage{amsthm}
\usepackage{amsmath}
\usepackage{amssymb}
\DeclareMathOperator{\lcm}{lcm}
\newtheorem{lemma}{Lemma}

\newtheorem{prop}{Proposition}
\newtheorem{thm}{Theorem}
\newtheorem{remark}{Remark}

\usepackage{version}
\usepackage[hidelinks]{hyperref}
\usepackage{authblk}
\usepackage{orcidlink}

\title{On the solving degree and complexity of Gr\"obner Basis computation over finite fields}

\author{Irene Di Muzio\,\orcidlink{0009-0007-2549-1087}
\thanks{\href{mailto:irene.muzio@uib.no}
{\texttt{irene.muzio@uib.no}}}}
\author{Igor Semaev%
\thanks{\href{mailto:igor.semaev@uib.no}
{\texttt{igor.semaev@uib.no}}}}
\affil{Department of Informatics, University of Bergen, Norway}

\date{}
\begin{document}

\maketitle

\begin{abstract}
Solving systems of multivariate polynomial equations over finite fields is a central problem in algebraic cryptanalysis, commonly approached using Gröbner basis algorithms. The complexity of Gröbner basis computation depends on the algorithm employed and is governed by the solving degree, defined as the maximum degree of the polynomials processed during the computation. Estimating this quantity is therefore crucial, and several bounds in terms of the degree of regularity have been established in the literature.

In this work, we  investigate the effect of Buchberger's criteria on the elimination of redundant pairs, and  prove  that the solving degree of a Gröbner basis algorithm combining linear algebra with Buchberger's original algorithm is bounded above by the degree of regularity defined by the leading forms of the involved polynomials. This result provides a rigorous justification for using the degree of regularity in complexity estimates for Gröbner basis algorithms, thereby supporting a common assumption in algebraic cryptanalysis.

We  refine the complexity analysis by establishing bounds on the number of pairs that must be processed. These bounds improve upon published complexity estimates, with particularly significant gains in certain parameter ranges.

\end{abstract}

\section{Introduction}\label{sec:intro}
The development of efficient methods for solving multivariate polynomial systems over finite fields is an important problem in computational algebra, with applications in  cryptography.
A classical approach to solving polynomial systems is Gr\"obner basis computation. After the first algorithm proposed by Buchberger~\cite{B65}, Lazard established a connection between Gr\"obner basis computation and Gaussian elimination on matrices constructed from polynomial multiples~\cite{L83}.
This connection is exploited by more efficient Gr\"obner basis algorithms, notably F4~\cite{F4}, F5~\cite{F5}, as well as the related method XL~\cite{XL}, and all their variants.

We study the complexity of constructing a Gr\"obner basis for the polynomial system over $\mathbb F_q$
\begin{equation}\label{eq:s1}
    f_1,\dots,f_m,\, x_1^q-x_1,\dots,x_n^q-x_n.
\end{equation}
Let $R^h=\mathbb F_q[x_1,\dots,x_n]/\langle x_1^q,\dots,x_n^q\rangle$.
Denote by $l_q(n,d)$ the number of monomials of degree $d$ in $R^h$, and by $L_q(n,D)=\sum_{d=0}^D l_q(n,d)$ the number of monomials of degree at most $D$.
Let $I^h$ be the ideal of $R^h$ generated by the leading forms of $f_1,\dots,f_m$, and let $V_d$ denote its subspace of homogeneous elements of degree $d$. The smallest positive integer $d$ such that $V_d$ contains every form of degree $d$ in $R^h$ is called the degree of regularity of the system~\eqref{eq:s1} and is denoted by $d_{\mathrm{reg}}$~\cite{BFS03}.
For $d=d_{\mathrm{reg}}$, the space $V_d$ is generated by all monomials of degree $d$ in $R^h$, and hence $\dim V_d=l_q(n,d)$.

Note that this definition of the degree of regularity takes the field equations into account, in particular through the definition of $R^h$. Consequently, the degree of regularity defined in this way is smaller than that of $f_1,\ldots,f_m$ considered without the field equations in $\mathbb{F}_q[x_1,\ldots,x_n]$, whenever the latter is well defined.
Nevertheless, our results also apply to polynomial systems without field equations, by carrying out the same construction in the ordinary polynomial ring and using the corresponding monomial counts.

Let $d_{\mathrm{solv}}$ denote the solving degree of the algorithm under consideration, the maximum degree of the polynomials processed during the computation.
The solving degree  governs the computational cost  of Gr\"obner basis algorithms, it  depends on the algorithm employed and is generally difficult to estimate directly.
Bounds on the solving degree are therefore often expressed in terms of the degree of regularity. The degree of regularity has been extensively studied, particularly for semi-regular polynomial systems~\cite{BFS04}.
In particular, the complexity estimates commonly used in algebraic cryptanalysis for F4 and F5 are of the form $O\bigl(\binom{n+d_{\mathrm{reg}}}{d_{\mathrm{reg}}}^{\omega}\bigr)$ and take the degree of regularity as a proxy for the solving degree. This is justified for homogeneous regular and semi-regular sequences~\cite{BFS04}, whereas for inhomogeneous systems it is in general an assumption. XL is analysed similarly, in terms of its own operating degree~\cite{XLdeg}, and aims at computing a solution rather than a Gr\"obner basis.

Several approaches have been developed to relate the solving degree to other parameters arising in polynomial system solving. 
Caminata and Gorla established bounds involving the Castelnuovo-Mumford regularity and studied the connections between the solving degree, the last fall degree, and the degree of regularity~\cite{CG21,CG23}.

In a different approach, Semaev and Tenti proved the bound $d_{\mathrm{solv}}\le 2d_{\mathrm{reg}}-2$ for a Buchberger-based algorithm over finite fields, without requiring semi-regularity or genericity~\cite{ST}.
Salizzoni subsequently established the bound $d_{\mathrm{solv}}\le d_{\mathrm{reg}}+1$, under suitable degree assumptions~\cite{S25}. 

In this work, we follow the approach of Semaev and Tenti by constructing an intermediate polynomial system at the degree of regularity. Exploiting the structure of this system, we show that Buchberger's second criterion can be relaxed: a critical pair can be discarded as soon as a third leading monomial divides its least common multiple, a condition that is not sufficient in general. This observation implies the stronger bound $d_{\mathrm{solv}}\le d_{\mathrm{reg}}$ for the resulting algorithm.
In~\cite[Example 3.9]{S25}, the bound $d_{\mathrm{solv}}\le d_{\mathrm{reg}}+1$ is shown to be sharp. However, in general, any implementation of Gr\"obner basis algorithms can compute S-polynomials by omitting the leading terms that cancel identically, as illustrated by~\eqref{eq:Spol}. In particular, in this example, this avoids explicitly constructing polynomial multiples of degree $d_{\mathrm{reg}}+1$.

Whereas the bound in~\cite{S25} is obtained by controlling the degrees of representations of the polynomials, our approach works directly at the level of critical pairs: Lemmas~\ref{lem:redundantNOfield} and~\ref{lem:redundantField} provide a transparent description of how critical pairs are handled by the algorithm.  This allows us not only to prove the degree bound, but also to identify the critical pairs that need to be processed and derive refined upper bounds on their number, as shown in Lemma~\ref{lem:N1N2}.

Kouba, Neiger, and Safey El Din also study the elimination of redundant critical pairs using a syzygy-based version of Buchberger's criteria in F4 Tracer~\cite{F4Tracer}. Their analysis concerns generic zero-dimensional systems and is subject to several assumptions, including conjectural properties of grevlex initial ideals. In contrast, our combinatorial analysis uses only the prescribed leading monomials of the intermediate generating set at the degree of regularity, without requiring genericity, conjectural properties, or any specific monomial order.

The improved degree bound in our work, together with the reduction in the number of pairs that need to be processed, leads to a significant improvement in the complexity of constructing a Gr\"obner basis.
Set $N=L_q(n,d_{\mathrm{reg}}).$ Theorem~\ref{thm:compl}  shows that, when $d_{\mathrm{reg}}=o(n)$  and $q$ is fixed, the computation of a Gr\"obner basis has complexity $O(N^{3+o(1)})$ operations in $\mathbb{F}_q$.
For the remaining case $d_{\mathrm{reg}}=\Omega(n)$, we establish a general complexity bound of $O(N^4)$, and, in certain parameter ranges, obtain an exponential improvement to $O(c^n N^4)$ for some constant $c<1$, as shown in Theorem~\ref{thm:complLargeD}. That is an improvement on the complexity of the Semaev-Tenti algorithm in~\cite{ST}.
We also recall in Section~\ref{sec:sol} the procedure of~\cite{ST} to compute a solution of the system or prove that none exists. Such procedure applies directly to the Gr\"obner basis constructed by our algorithm and requires $O(nN^3)$ additional operation for fixed $q.$

In~\cite{S25} the complexity bound  $O(M^4+Mk^{\omega-1})$ for computing  a Gr\"obner basis was proved, where $k$ is the number of input polynomials and $M=\binom{n+d_{\mathrm{reg}}+1}{d_{\mathrm{reg}}+1}$.
Compared with these estimates, our bounds benefit from a smaller monomial count than in~\cite{S25}, as $N\le\binom{n+d_{\mathrm{reg}}}{d_{\mathrm{reg}}}<M$  and   achieve a significantly  lower asymptotic complexity in certain parameter ranges.

The complexity estimates for F4, F5 and XL recalled above rely on an unproved assumption that the solving degree in these algorithms do not exceed the degree of regularity. That fact is rigorously proved for the algorithm in the present work.  Our approach was tailored for finding zeroes to polynomial systems in the ground finite field. So, the definition of the degree of regularity depends on the field polynomials thus providing generally smaller numerical values for the degree of regularity and the complexity than in general.

The complexity bounds in this work depend only on the leading forms of the polynomials and not on their lower degree terms. For the present algorithm, the bound $d_{\mathrm{solv}}\le d_{\mathrm{reg}}$ is in fact an equality, since the intermediate generating set is constructed at degree $d_{\mathrm{reg}}$ from the leading forms alone, ignoring the lower degree relations that arise when the leading forms of combinations of the input polynomials cancel. By exploiting them,  a Gr\"obner basis may be reached in the same or a strictly smaller degree, with $d_{\mathrm{reg}}$ remaining an upper bound.

In developing the present work, Di Muzio initially studied the Gebauer-M\"oller implementation of Buchberger's algorithm~\cite{GM}, and provided a combinatorial proof of the bound $d_{\mathrm{solv}}\le d_{\mathrm{reg}}+1$ for this implementation on a class of systems over $\mathbb F_2$.
Semaev subsequently observed that incorporating the criteria for eliminating redundant pairs into the Semaev-Tenti algorithm allows the bound to be improved to $d_{\mathrm{reg}}$, and proved this result over $\mathbb F_2$.
This approach also allowed for straightforward complexity estimates in terms of the degree of regularity.
Di Muzio then generalized the result to arbitrary finite fields $\mathbb F_q$, as presented in Section~\ref{sec:solvD}, and carried out the complexity analysis in Section~\ref{sec:compl}.

\section{The improved solving degree bound}\label{sec:solvD}
We first recall the construction of Semaev and Tenti in Section \ref{sec:ST}, which provides a generating system with a prescribed structure of leading terms.
Then,  in Section \ref{sec:criteria}, we show how Buchberger's criteria for eliminating redundant pairs can be applied to such a system to bound the solving degree by the degree of regularity.
Finally, in Section \ref{sec:sol}, we recall how a solution, if exists, can be computed once a Gr\"obner basis has been constructed.

\subsection{Preliminary construction of the polynomial system}\label{sec:ST}
Let $R=\mathbb F_q[x_1,\dots,x_n]$ be the polynomial ring in $n$ variables over the finite field $\mathbb F_q$.
For an $n$-tuple of non-negative integers $\alpha=(\alpha_1,\dots,\alpha_n)$, denote by $x^\alpha=x_1^{\alpha_1}\cdots x_n^{\alpha_n}$ the corresponding monomial, of total degree $|\alpha|=\sum_{i=1}^n\alpha_i$.
Fix a total-degree monomial order on $R$.

Let $I$ be the ideal in $R$ generated by the polynomials in \eqref{eq:s1}.
As shown in \cite{ST}, one can construct a intermediate generating set $B$ of $I$ with the following two properties.
First, every polynomial in $B$ has degree at most $d_{\mathrm{reg}}$. 
Second, there are exactly $l_q(n,d_{\mathrm{reg}})$ polynomials whose leading monomials range over all monomials of degree $d_{\mathrm{reg}}$ in $R^h$, and whose remaining terms have strictly smaller degree.

When $q>d_{\mathrm{reg}}$, the field polynomials $x_i^q-x_i$ can be reduced to polynomials of degree less than $d_{\mathrm{reg}}$. When $q \le d_{\mathrm{reg}}$, the basis $B$ explicitly includes the field polynomials. That is
\begin{equation}\label{eq:s2}
B=\{f_\rho, f_\sigma\}_{\rho,\sigma}\cup 
\begin{cases}
\{x_1^q-x_1,\ldots,x_n^q-x_n\}, & q\le d_{\mathrm{reg}},\\
\varnothing, & q>d_{\mathrm{reg}}.
\end{cases}
\end{equation}
Every monomial term $x^{\alpha}$ appearing in the polynomials in $\{f_\rho, f_\sigma\}_{\rho,\sigma}$, with $\alpha=(\alpha_1,\ldots,\alpha_n)$, satisfies $0\le\alpha_i<q$ for all $i$ and 
\begin{itemize}
    \item $f_\rho =x^\rho  + g_\rho $, where $x^\rho,\, |\rho |=d_{\mathrm{reg}}$ is the leading monomial  of $f_\rho$ and $\deg(g_\rho )<|\rho |$;
    \item $f_\sigma=x^\sigma+g_\sigma$, where $x^\sigma,\,|\sigma|<d_{\mathrm{reg}}$ is the leading monomial of $f_\sigma$ and $\deg(g_\sigma)\le|\sigma|$.
\end{itemize}

\subsection{Application of Buchberger's criteria}\label{sec:criteria}
For each pair of polynomials $f_\alpha,f_\beta$ in the current basis, Buchberger's algorithm computes the $S$-polynomial $S(f_{\alpha},f_{\beta})$ and reduces it. Every non-zero remainders is added to the basis. The algorithm terminates when all such reductions are zero.
Applying Buchberger's algorithm to \eqref{eq:s2}, when either $q>d_{\mathrm{reg}}$ or $q\le d_{\mathrm{reg}}$,  yields a basis of the same form.

For every pair of polynomials $f_\alpha,f_\beta$ in the basis, define the monomial $x^\delta=\lcm(x^\alpha,x^\beta)$, where $\delta=\delta(\alpha,\beta)=(\delta_1,\dots,\delta_n)$ and $\delta_i=\max(\alpha_i,\beta_i).$

Buchberger's criteria allow certain pairs $f_\alpha,f_\beta$  to be discarded during the computation~\cite{B79, B85} by the following rules.
\begin{itemize}
    \item[(C1)] The leading monomials $x^\alpha$ and $x^\beta$ are coprime.
    \item[(C2)] There exists a third leading monomial $x^\gamma$ dividing $x^\delta$, such that each of the pairs $(f_\alpha,f_\gamma)$ and $(f_\gamma,f_\beta)$ is dealt with, either by processing it  by reducing the $S$-polynomial or by discarding it  with Buchberger's criteria.% or has an lcm of degree strictly smaller than $|\delta|$.
\end{itemize}
In either case, the pair $f_\alpha,f_\beta$ is said to be redundant, and there is no need to reduce $S(f_{\alpha},f_{\beta})$.

The next lemmas investigates the combinatorial structure of Buchberger's second criterion (C2) within the specific basis \eqref{eq:s2}, and turns it into a direct strategy for detecting redundant pairs from the exponent tuples.

\begin{lemma}\label{lem:redundantNOfield}
Consider a pair $f_\alpha,f_\beta$, with $\alpha\neq\beta$, from the first set in \eqref{eq:s2}. Assume one of the following conditions:
\begin{itemize}
    \item[(i)] $|\alpha|=|\beta|=d_{\mathrm{reg}}$ and $|\delta|\ge d_{\mathrm{reg}}+2$;
    \item[(ii)] $|\alpha|<d_{\mathrm{reg}}  $ or $ |\beta|<d_{\mathrm{reg}}$, and $|\delta|\ge d_{\mathrm{reg}}+1$.
\end{itemize}
Then there exists an $n$-tuple $\gamma\neq\alpha,\beta$ such that 
\begin{enumerate}
\item $|\gamma|=d_{\mathrm{reg}}$ and $x^\gamma \mid x^\delta$,
\item $\delta(\alpha,\gamma)=d_{\mathrm{reg}}+1$ and\quad $\delta(\beta,\gamma)=\delta(\alpha,\beta)-1$. The latter holds if  $|\beta|<d_{\mathrm{reg}}$ in case (ii).  
\end{enumerate}

\end{lemma}
\begin{proof}
Define
\[
K=\{k\in\{1,\dots,n\}:\alpha_k<\beta_k\},
\]
and choose $j$ such that $\alpha_j>\beta_j$.
Such indices $K$ and $j$  exist; otherwise one monomial would divide the other, which would imply $|\delta|\le d_{\mathrm{reg}}$, contradicting the assumptions.

\noindent\textbf{Case (i).} Choose any $k\in K$, and define $\gamma$ by setting
\[
\gamma_k=\alpha_k+1,\qquad \gamma_j=\alpha_j-1,
\]
while $\gamma_i=\alpha_i$ for every $i \ne j,k$. This gives $|\gamma|=|\alpha|=d_{\mathrm{reg}}$, and $\gamma_i\le\delta_i$ for any $i$, so $x^\gamma\mid x^\delta$. Clearly, $\gamma\neq\alpha$. The assumption $|\delta| \ge d_{\mathrm{reg}}+2$ ensures $\gamma\ne\beta$. That proves the first statement of the Lemma. The second statement obviously follows.

\noindent\textbf{Case (ii).} By symmetry, we may assume $|\beta|<d_{\mathrm{reg}}$.
Choose integers $\gamma_k$ for $k\in K$ such that
\begin{equation}\label{gamma_k}
\alpha_k\le\gamma_k\le\beta_k,
\qquad
\sum_{k\in K}(\gamma_k-\alpha_k)=d_{\mathrm{reg}}-|\alpha|+1.
\end{equation}
Set $\gamma_j=\alpha_j-1$, and set  $\gamma_i=\alpha_i$ for $i\notin K$ and $i\ne j$. 
Then
\[
|\gamma|=\sum_{k \in K} \gamma_k+ \sum_{i \notin K} \alpha_i-1 =\sum_{k\in K}(\gamma_k-\alpha_k)+|\alpha|-1=d_{\mathrm{reg}},
\]
and, componentwise, $\gamma_i\le\delta_i$.
The existence of such  $\gamma_k$ follows from
\[
|\delta|=|\alpha|+\sum_{k\in K}(\beta_k-\alpha_k)\ge d_{\mathrm{reg}}+1,
\]
that holds by assumption on $|\delta|$.
Hence, $\sum_K (\beta_k - \alpha_k) \ge d_{\mathrm{reg}}-|\alpha|+1$, where $\beta_k - \alpha_k \ge 0$. So, one may choose $\gamma_k$ such that  \eqref{gamma_k} holds. 

Moreover, $\gamma\neq\alpha$ by construction, and $\gamma\neq\beta$ because $|\gamma|=d_{\mathrm{reg}}$ whereas $|\beta|<d_{\mathrm{reg}}$. That proves the first statement of the Lemma. The second statement obviously follows.
\end{proof}

\begin{lemma}\label{lem:redundantField}
Let $f_\alpha=x_j^q-x_j$ be a field polynomial in \eqref{eq:s2}, and let $f_\beta$ belong to the first set in \eqref{eq:s2}, with $\beta_j>0$. Assume one of the following conditions:
\begin{itemize}
    \item[(i)] $|\beta|=d_{\mathrm{reg}}$
    and $|\delta|\ge d_{\mathrm{reg}}+2$,
    \item[(ii)] $|\beta|<d_{\mathrm{reg}}$
    and $|\delta|\ge d_{\mathrm{reg}}+1$.
\end{itemize}
Then there exists an $n$-tuple $\gamma=(\gamma_1,\ldots,\gamma_n), \,0\le\gamma_i<q$ and $\gamma\neq\alpha,\beta$ such that 
\begin{enumerate}
\item $|\gamma|=d_{\mathrm{reg}}$ and $x^\gamma \mid x^\delta$,
\item $\delta(\alpha,\gamma)=d_{\mathrm{reg}}+1$ and\quad $\delta(\beta,\gamma)=\delta(\alpha,\beta)-1$.   
\end{enumerate}

\end{lemma}
\begin{proof}
We use a construction analogous to that in the previous proof. Indeed, $|\delta|=q+\sum_{k\ne j}\beta_k\geq d_{\mathrm{reg}}+1$  holds in both (i) and (ii). So, $\sum_{k\ne j}\beta_k\geq d_{\mathrm{reg}}-q+1$.

Set $\gamma_j=q-1$ and choose the remaining coordinates of $\gamma$ so that 
\[
0\le \gamma_k\le\beta_k \text{ for }k\neq j, \qquad \sum_{k\ne j}\gamma_k=d_{\mathrm{reg}}-q+1.
\]
Then $|\gamma|=d_{\mathrm{reg}}$, each coordinate satisfies $0\le\gamma_i<q$ and $x^\gamma\mid x^\delta$.
Clearly, $\gamma\ne\alpha$.

In case (i), $|\delta|\ge d_{\mathrm{reg}}+2$ implies $\beta_j\le q-2$, hence $\gamma_j=q-1>\beta_j$ and $\gamma\ne\beta$.
In case (ii), $\gamma\ne\beta$ follows from $|\gamma|=d_{\mathrm{reg}}>|\beta|$. That proves the first statement of the Lemma. The second statement obviously follows.
\end{proof}

\begin{remark}
The purpose of the constructions in Lemmas~\ref{lem:redundantNOfield} and~\ref{lem:redundantField} is to find a polynomial $f_\gamma$ in \eqref{eq:s2}.
Thus, the bounds on $|\delta|$ cannot be improved further.
Indeed, the assumptions ensure that one can find a tuple $\gamma$ of degree $d_{\mathrm{reg}}$ and, by construction, $x^\gamma$ is the leading monomial of some polynomial in \eqref{eq:s2}.
For smaller values of $|\delta|$, one may construct tuples $\gamma$ of degree $<d_{\mathrm{reg}}$, but then there is no guarantee that $x^\gamma$ occurs as a leading monomial in the system.
\end{remark}

As observed by Buchberger~\cite{B79}, the existence of a third leading monomial dividing the least common multiple is not, in general, sufficient to discard a critical pair. However, the particular structure of the generating set~\eqref{eq:s2} allows us to show that the auxiliary pairs can be recursively reduced to the pairs selected by our algorithm, as we show in the following.

\begin{prop}\label{prop:redundant}
The pairs $(f_\alpha,f_\beta)$ satisfying conditions (i) or (ii) of Lemma~\ref{lem:redundantNOfield} or Lemma~\ref{lem:redundantField} are redundant.
\end{prop}
\begin{proof}
Consider such a pair $(f_\alpha,f_\beta)$ and let $\delta=\delta(\alpha,\beta)$. By the preceding lemmas, there exists a third tuple $\gamma$ such that $|\gamma|=d_{\mathrm{reg}}$ and $x^{\gamma}|x^{\delta}$. In all cases we have
\begin{equation}\label{equalities}
|\delta(\gamma,\alpha)|=d_{\mathrm{reg}}+1,
\qquad
|\delta(\gamma,\beta)|=\delta(\alpha,\beta)-1.
\end{equation}
In what follows, we show that  $(f_\alpha,f_\beta)$ may be discarded  by criterion (C2). We proceed by induction on $|\delta(\alpha,\beta)|$ to show that auxiliary pairs $(f_\gamma,f_\alpha)$ and $(f_\gamma,f_\beta)$ do not satisfy the conditions of the lemmas (and should be processed) or can themselves be discarded using criterion (C2), possibly through further auxiliary pairs.
We consider two cases.
\begin{enumerate}
\item  Let $|\delta(\alpha,\beta)|=d_{\mathrm{reg}}+1$. Assume first that neither of $f_\alpha,f_\beta$ is a field polynomial. If $|\alpha|=|\beta|=d_{\mathrm{reg}}$, then the pair does not satisfy the condition (i) of Lemma~\ref{lem:redundantNOfield}.  

So, at least one of $|\alpha|,|\beta|$ must be smaller than $d_{\mathrm{reg}}$. By symmetry,  $|\beta|<d_{\mathrm{reg}}$.
Since $\delta(\gamma,\beta)=d_{\mathrm{reg}}$ by \eqref{equalities}, the pair $(f_\gamma,f_\beta)$ does not satisfy the conditions of the lemmas.

Let's  consider $(f_\gamma,f_\alpha)$. We have $|\gamma|=d_{\mathrm{reg}}$ and $|\delta(\gamma,\alpha)|=d_{\mathrm{reg}}+1$.
If $|\alpha|=d_{\mathrm{reg}}$, then the leading monomials of  $(f_\gamma,f_\alpha)$  have degree $d_{\mathrm{reg}}$ and  $|\delta(\gamma,\alpha)|=d_{\mathrm{reg}}+1$. So $(f_\gamma,f_\alpha)$  does not satisfy the conditions of the lemmas.

Let $|\alpha|<d_{\mathrm{reg}}$, applying the construction of Lemma~\ref{lem:redundantNOfield} to $(f_\gamma,f_\alpha)$, we obtain a tuple $\gamma'$, where $|\gamma'|=d_{\mathrm{reg}}$ and
\[
\delta(\gamma,\gamma')=d_{\mathrm{reg}}+1,
\qquad
\delta(\gamma',\alpha)=d_{\mathrm{reg}}.
\]
Both auxiliary pairs $(f_{\gamma'},f_\gamma)$ and $(f_{\gamma'},f_\alpha)$ do not satisfy the conditions of the lemmas. Thus, (C2) allows  to discard $(f_\gamma,f_\alpha)$, and consequently $(f_\alpha,f_\beta)$.

Assume now $f_\alpha=x_j^q-x_j$ is a field polynomial. Since $|\delta(\alpha,\beta)|=d_{\mathrm{reg}}+1$, condition (i) of Lemma~\ref{lem:redundantField} cannot hold. 
Thus the pair must satisfy condition (ii), and hence $|\beta|<d_{\mathrm{reg}}$. The construction of $\gamma$ in Lemma~\ref{lem:redundantField} gives $\gamma_j=q-1$, and $|\gamma|=d_{\mathrm{reg}}$, and $\delta(\gamma,\alpha)=d_{\mathrm{reg}}+1$.
Hence $(f_\gamma,f_\alpha)$ does not satisfy the conditions of the lemmas. Moreover, $\delta(\gamma,\beta)=d_{\mathrm{reg}}$ by \eqref{equalities}. So, $(f_\gamma,f_\beta)$ does not satisfy the conditions of the lemmas.
\item Let $|\delta(\alpha,\beta)|\geq d_{\mathrm{reg}}+2$. By  \eqref{equalities}, we have
\[
\delta(\gamma,\alpha)<|\delta(\alpha,\beta)|,
\qquad
\delta(\gamma,\beta)<|\delta(\alpha,\beta)|,
\]
Thus both auxiliary pairs $(f_\gamma,f_\alpha)$ and $(f_\gamma,f_\beta)$ have least common multiples of strictly smaller degree. By the induction hypothesis, each auxiliary pair does not satisfy the conditions of the lemmas or can itself be discarded by criterion (C2). Consequently, criterion (C2) also applies to $(f_\alpha,f_\beta)$.

\end{enumerate}
\end{proof}

\subsection{The new algorithm}\label{sec:alg}

For an $n$-tuple $\alpha$, let $\operatorname{supp}(\alpha)=\{i:\ \alpha_i>0\}$.
\begin{enumerate}
\item Starting from \eqref{eq:s1}, construct the basis $B$ as in \eqref{eq:s2}, according to \cite{ST}.
\item  Let $f_\alpha\ne f_\beta$ belong to the current  basis and $\hbox{supp}(\alpha)\cap \hbox{supp}(\beta)\ne\varnothing$. Assume one of 
\begin{itemize}
    \item  $|\alpha|=|\beta|=d_{\mathrm{reg}}$ and $|\delta(\alpha,\beta)|=d_{\mathrm{reg}}+1$,
    \item $|\alpha|$ or $|\beta|<d_{\mathrm{reg}}$ and $|\delta(\alpha,\beta)|\leq d_{\mathrm{reg}}$,
    \item $f_{\alpha}=x_j^q-x$, and $|\beta|=d_{\mathrm{reg}}$, and $|\delta(\alpha,\beta)|=d_{\mathrm{reg}}+1$,
    \item $f_{\alpha}=x_j^q-x$, and $|\beta|<d_{\mathrm{reg}}$, and $|\delta(\alpha,\beta)|\leq d_{\mathrm{reg}}$.  
\end{itemize}
Then reduce $S(f_\alpha,f_\beta)$  modulo the current basis. Add the remainder to the basis and repeat.
\item Terminate if all the remainders are zero.
\end{enumerate}
The system  \eqref{eq:s2} contains only polynomials with leading monomials of degree $\le d_{\mathrm{reg}}$.  Every time the residue of an S-polynomial is non-zero, its degree is $<d_{\mathrm{reg}}$ and it is added to the basis. We can assume that the polynomials in the basis are linearly independent at every step of the algorithm.
Throughout the whole algorithm, the basis is made of at most $N=L_q(n,d_{\mathrm{reg}})$ polynomials if $q > d_{\mathrm{reg}}$, or  at most $N + n$ polynomials when $q \le d_{\mathrm{reg}}$.

This algorithm constructs a Gr\"{o}bner basis without ever producing polynomials of degree higher than $d_{\mathrm{reg}}$.
Indeed, rewriting an $S$-polynomial as
\begin{align}\label{eq:Spol}
   S_{\alpha,\beta}=S(f_\alpha,f_\beta)=\dfrac{x^\delta}{x^\alpha}f_\alpha - \dfrac{x^\delta}{x^\beta}f_\beta 
   =\dfrac{x^\delta}{x^\alpha}g_\alpha - \dfrac{x^\delta}{x^\beta}g_\beta,
\end{align}
observe that $\deg(S_{\alpha,\beta}) \le \max\{|\delta|-|\alpha|+\deg(g_\alpha), |\delta|-|\beta|+\deg(g_\beta)\}$. By the construction in \eqref{eq:s2}, $\deg(g_\rho)<|\rho|$ if $|\rho|=d_{\mathrm{reg}}$, whereas $\deg(g_\sigma) \le |\sigma|$ if $|\sigma|<d_{\mathrm{reg}}$. Hence,
\[\deg(S_{\alpha,\beta})\le
\begin{cases}
 |\delta| -1, &\text{ if } |\alpha|=|\beta|=d_{\mathrm{reg}},\\
 |\delta|, &\text{ if } |\alpha|<d_{\mathrm{reg}}  \text{ or } |\beta|<d_{\mathrm{reg}}.
\end{cases}
\]
This implies  $\deg(S_{\alpha,\beta})\le d_{\mathrm{reg}}$ for all cases except $f_{\alpha}=x_j^q-x$, and $q<|\beta|=d_{\mathrm{reg}}$, and $|\delta(\alpha,\beta)|=d_{\mathrm{reg}}+1$. However, in this case $\beta_j=q-1$ and
\[
S_{\alpha,\beta}=-x_j\prod_{i\ne j}x_i^{\beta_i}-x_jg_\beta.
\]
Since $\deg(g_\beta)<d_{\mathrm{reg}}$ and $q\ge 2$, we obtain
\[
\deg(S_{\alpha,\beta})\le
\max\{d_{\mathrm{reg}}-q+2,\,
1+\deg(g_\beta)\}
\le d_{\mathrm{reg}}.
\]
Hence, every S-polynomial selected by the algorithm has degree at most $d_{\mathrm{reg}}$.

Thanks to the modifications above,  the algorithm reduces only polynomials of degree  $\le d_{\mathrm{reg}}$, and the remainders that are added have degree  $<d_{\mathrm{reg}}$. After each iteration, the system keeps the structure \eqref{eq:s2}.
The algorithm terminates with solving degree $\le d_{\mathrm{reg}}.$

\subsection{Computing a solution}\label{sec:sol}
Once a Gr\"obner basis has been constructed, a solution to the system defined by the polynomials in \eqref{eq:s1}, or a proof that no solution exists, can be obtained by the procedure described in~\cite[Theorem 2.2]{ST}. Convert the basis into a reduced Gröbner basis in $O(N^3)$ field operations, by removing redundant leading monomials and reducing the remaining polynomials. If the reduced basis is $\{1\}$ then there are no solutions; otherwise iteratively assign values $a_i\in\mathbb F_q$ to the variables by adjoining the linear polynomials $x_i-a_i$ and updating the Gr\"obner basis, discarding assignments that yield the basis $\{1\}$. Each update requires $O(N^3)$ field operations. Since at most $qn$ assignments are tested, the additional cost is $O(qnN^3)$, which reduces to $O(nN^3)$ when $q$ is fixed.
For fixed $q$, the cost of computing a solution, when one exists, is absorbed by the complexity bounds for constructing a Gr\"{o}bner basis established in Theorem~\ref{thm:compl}.

\section{Complexity of constructing  Gr\"{o}bner bases}\label{sec:compl}
In this section we estimate the complexity of constructing  a Gr\"{o}bner basis from the polynomial system \eqref{eq:s2} according to the algorithm  in Section \ref{sec:solvD}.
We compare the resulting bounds with those obtained for the algorithm of Semaev and Tenti \cite{ST}.
Our improvement  is  due to both a lower solving degree and the reduced number of pairs to be processed, by the application of Buchberger criteria. 

Assume that all parameters  $q$, $d_{\mathrm{reg}}$ and $n$ may grow. Suppose $d_{\mathrm{reg}}\ge 2$. Note that $d_{\mathrm{reg}}$ is subject to the constraint $d_{\mathrm{reg}}\le n(q-1)$. Indeed, every monomial in $R^h$ has exponent at most $q-1$ in each variable, and hence its degree is at most $n(q-1)$. 
We also assume that every polynomial $f_i$ in \eqref{eq:s1} has degree at most $d_{\mathrm{reg}}$. Otherwise, each $f_i$ with $\deg(f_i)>d_{\mathrm{reg}}$ is first replaced by its remainder after division by the polynomials in \eqref{eq:s2}, at an additional cost of $O\big(L_q(n,\deg(f_i))\,L_q(n,d_{\mathrm{reg}})\big)$.
For ease of notation, we set $D=d_{\mathrm{reg}}$  and $N=L_q(n,D)$ throughout this whole section.

The algorithm presented in \cite{ST} consists of two main stages. The first determines the degree of regularity and constructs the basis \eqref{eq:s2} by means of Macaulay triangulations, together with the necessary reductions of the field polynomials. The complexity of this stage was bounded by $O(D^2N^4+qN^2)$ field operations. The second stage applies the plain Buchberger's algorithm to the polynomials in \eqref{eq:s2} to construct the Gr\"{o}bner basis. Its complexity was bounded by $O(N^6)$, and can be reduced to $O(N^5)$ by keeping track of the already processed $S$-polynomials. We use the latter bound for the comparison below.

For the present algorithm, every $S$-polynomial to be processed has degree  $\le D$, according to Section \ref{sec:solvD}, and it is divided by the polynomials in the current basis, whose degrees  $\le D$ as well. Thus a single reduction requires at most $O(N^2)$ field operations. Moreover, the algorithm does not have to process all $\binom{N}{2}= O(N^2)$ possible pairs. By Lemma~\ref{lem:redundantNOfield}, it is sufficient to process at most $N_1+N_2$ pairs not involving field polynomials.
By $N_1$ we denote  the number of unordered pairs $f_\alpha\ne f_\beta$ in \eqref{eq:s2}, where supp$(\alpha)\cap$ supp$(\beta)\ne \varnothing$, and such that $|\alpha|=|\beta|=D$, and $|\delta|=D+1.$ By $N_2$ we denote the number of unordered pairs $f_\alpha\ne f_\beta$, where supp$(\alpha)\cap$ supp$(\beta)\ne \varnothing$, and such that $|\delta| \le D.$ 

In addition, pairs involving a field polynomial need to be considered only when $q\le D$. Since there are $n$ field polynomials and each can be paired with at most $N$ polynomials from $\{f_{\rho},f_\sigma\}_{\rho,\sigma}$, their total number is bounded by $\mathbf{1}_{\{q\le D\}}nN$. This coarse bound suffices, since its contribution is absorbed into the asymptotic bounds derived below.

Let
\[
\rho = \frac{N_1+N_2+\mathbf{1}_{\{q\le D\}}nN}{N^2},
\]
which depends on $n,D,q$, so that the second stage complexity becomes $O\left(\rho N^4\right).$
We show in Theorem~\ref{thm:compl} that in the case $D=o(n)$, the decay of $\rho$ yields a further improvement, approaching two powers of $N$ compared with the previous $O(N^5)$ bound from~\cite{ST}.

We first estimate $N_1$ and $N_2$ in Section~\ref{sec:numPairs}.
In Section~\ref{sec:newCompl}, we combine these bounds with refined costs of the first stage, to analyze the complexity in different cases, and show the improvement over the previous bound. Finally, we provide the memory complexity of the complete algorithm in Section~\ref{sec:mem}.

\subsection{The number of processed pairs}\label{sec:numPairs}

Let $l_c(n,d)=|\{ (\alpha_1,\dots, \alpha_n): \sum_i \alpha_i =d,\ 0\le \alpha_i <c\}|$ for any $c\geq 1$.
To simplify the notation in what follows, we assume $\alpha=(\alpha_1,\dots,\alpha_n),\, 0\le \alpha_i<q$ and $\beta=(\beta_1,\dots,\beta_n),\, 0\le \beta_i<q$. So, $\delta=\delta(\alpha,\beta)=(\delta_1,\dots,\delta_n)$ and $\delta_i=\max(\alpha_i,\beta_i)$. Moreover, set $\gamma=\gamma(\alpha,\beta)=(\gamma_1,\dots,\gamma_n)$ where $\gamma_i=\min(\alpha_i,\beta_i)$. Finally, set $B_q(z)=\sum_{j=0}^{q-1}(2j+1)z^j.$

The following lemma gives bounds on the number of processed pairs, according to the redundancy result of the previous section. As explained, pairs involving a field polynomial are included whenever $q\le D$.
The bounds used below are sufficient for the complexity analysis. The corresponding exact counting formulas, together with a refined buond for pairs involving a field polynomial, are given in Appendix~\ref{app:exactPairs}.

\begin{lemma}\label{lem:N1N2}
\begin{equation*} N_1\leq 
\binom{n}{2}l_q(n,D-1),
\qquad
N_2\leq\frac{1}{2} \sum_{t=1}^{D}[z^t]B_q(z)^n. \label{eq:boundPair}
\end{equation*}
\end{lemma}

\begin{proof}
We first count the pairs contributing to $N_1$ which do not involve field polynomials. Since $|\alpha|=|\beta|=D$ and $|\delta|=D+1,$ the trivial identity $|\gamma|+|\delta|=|\alpha|+|\beta|$ gives $|\gamma|=D-1.$
Therefore, there exist two distinct indices $i,j$ such that %$\alpha_i=\gamma_i+1,\ \beta_j=\gamma_j+1$.
$\alpha=\gamma+e_i$ and $\beta=\gamma+e_j$.
For each exponent tuple $\gamma$ of degree $D-1$, there are at most $\binom n2$ choices of distinct coordinates. Therefore we use the bound
\[
\binom n2 l_q(n,D-1).
\]

We now count the pairs contributing to $N_2$ not involving field polynomials. Fix $\delta.$
For one coordinate, the number of ordered pairs $(\alpha_i,\beta_i)$ satisfying $\max(\alpha_i,\beta_i)=\delta_i $ is $(\delta_i+1)^2-\delta_i^2=2\delta_i+1.$
Hence the number of ordered pairs $(\alpha,\beta)$ corresponding to such $\delta$ is $\prod_{i=1}^{n}(2\delta_i+1).$
Expanding $B_q(z)^n$ gives
\[
[z^t]B_q(z)^n=
\sum_{\substack{
|\delta|=t\\
0\leq\delta_i<q
}}
\prod_{i=1}^{n}(2\delta_i+1),
\]
so the number of ordered pairs with $0<|\delta|\leq D$ is
\[
\sum_{t=1}^{D}[z^t]B_q(z)^n.
\]
\end{proof}

\subsection{Improvements on the complexity}
\label{sec:newCompl}

Recall that the second stage has complexity $O(\rho N^4)$.
Let $T(n,D,q)$ denote the cost of the Macaulay triangulations in the first stage, and let $R(n,D,q)$ denote the additional cost of reducing the field polynomials when $q>D$.
The total complexity of the present algorithm is therefore bounded by
\begin{equation}
\label{total_complexity}
O\left(
T(n,D,q)+R(n,D,q)+\rho N^4
\right).
\end{equation}
We now derive refined bounds for $T(n,D,q)$ and $R(n,D,q)$.

The first stage determines the degree of regularity $D$ by successively constructing and triangulating, for $d\le D$, the Macaulay matrices of the leading forms of the polynomials of \eqref{eq:s1} in $R^h$. At degree $d$, the rows are obtained by multiplying each form of degree $k\le d$ by the monomials of degree $d-k$. At degree $D$, the row reduction produces leading forms covering all monomials of degree $D$. Keeping track of the same linear combinations of the original polynomials gives the polynomials used to construct the basis \eqref{eq:s2}. Thus the main computation in this stage is the triangulation of these Macaulay matrices.

The bound $O(D^2N^4)$ quoted above from \cite{ST} is obtained by using uniform upper bounds for their numbers of rows and columns. For the comparison with the new second stage complexity, it is useful to keep their degree-wise dimensions more precisely. 

Let $r_d$ denote the number of rows of the Macaulay matrix at degree $d\le D$. By the linear independence assumption, there are at most $l_q(n,k)$ input polynomials of degree $k$, each multiplied by the $l_q(n,d-k)$ monomials of degree $d-k$. Hence
\[
r_d\leq
\sum_{k=1}^{d}
l_q(n,k)l_q(n,d-k)
\leq l_q(2n,d).
\]
Since the degree-$d$ Macaulay matrix has $c_d=l_q(n,d)$ columns, its triangulation requires
\[
O\left(
l_q(2n,d)l_q(n,d)^2
\right)
\]
field operations. 
At degree $D$, the polynomials $f_\rho$ in \eqref{eq:s2} are constructed by selecting $c_D$ linearly independent rows and applying the inverse of the resulting $c_D\times c_D$  matrix to the corresponding inhomogeneous polynomials. This requires $O(c_D^2N)$ additional field operations. Since $N\le l_q(2n,D)$, this cost is absorbed by the degree-$D$ triangulation bound.
Summing over all degrees gives
\begin{equation}
\label{Mac_complexity}
T(n,D,q)=
O\left(
\sum_{d=0}^{D}
l_q(2n,d)l_q(n,d)^2
\right)=
O\left(
L_q(2n,D)L_q(n,D)^2
\right).
\end{equation}

When $q>D$, the $n$ field polynomials must also be reduced with respect to the computed basis. 
During a division the leading monomials that occur are of the form $x_i^b x^a$, with $0\le b\le q$ and $x^\alpha$ a monomial of degree $<D$ in the remaining $n-1$ variables, so there are at most $(q+1)L_q(n-1,D-1)$; each reduction step involves a basis polynomial $f_\rho$, which contains at most $L_q(n,D-1)+1 $ monomials. Thus, all the reductions require at most $O\left(nqL_q(n,D-1)^2\right)$ operations. Since $q>D$, the exponent cap is inactive up to degree $D$, and therefore
\[
N=
\binom{n+D}{D}
,
\quad
L_q(n,D-1)
=
\binom{n+D-1}{D-1}=
\frac{D}{n+D} N
.
\]
It follows that
\begin{equation}
\label{Red_complexity}
R(n,D,q)=
O\left(
\mathbf 1_{\{q>D\}}
qn\left(\frac{D}{n+D}\right)^2N^2
\right).
\end{equation}

In Theorem~\ref{thm:compl}, we show that the refined Macaulay contribution $T(n,D,q)$ is absorbed in the bounds for $\rho N^4$. However, the additional field polynomial reduction cost $R(n,D,q)$ must still be taken into account for the total complexity.
In particular, when $q$ grows, this contribution may dominate. On the other hand, assuming $q$ fixed, the cost  $R(n,D,q)$ is negligible compared to the cost of the other steps.

We now estimate these contributions in the two cases $D=o(n)$ and $D=\Omega(n)$, where the former means that $D/n\to 0$, and the latter means that $D\ge cn$ for some constant $c>0$ and all sufficiently large $n$. In the former, we obtain a sharper complexity bound, while in the latter we give a more general estimate. A refined analysis of the case $D=\Omega(n)$, yielding further improvements in certain ranges, is provided in Appendix~\ref{app:largeD}.

\begin{thm}\label{thm:compl}
The total complexity of the algorithm satisfies the following bounds.

\noindent\textup{(i)} If $D=o(n)$ and $D\to \infty$:
\[
O\left(N^{3+o(1)}+ \mathbf 1_{\{q>D\}}q \frac{D^2}{n}\,N^2 \right).
\]
\noindent\textup{(ii)} If $D=O(1)$:
\[
O(n^{3D+1}+\mathbf 1_{\{q>D\}}qn^{2D-1}).
\]

\smallskip
\noindent\textup{(iii)} If $D=\Omega(n)$:
\[
O(N^4+\mathbf 1_{\{q>D\}}qn\,N^2).
\]
\end{thm}

\begin{proof}
Recall the total complexity bound $O\left(T(n,D,q)+R(n,D,q)+\rho N^4\right).$
We estimate its three contributions separately in the two cases. We first estimate $\rho=(N_1+N_2+\mathbf{1}_{\{q\le D\}}nN)/N^2$ by bounding $N_1$ and $ N_2$, computed in Lemma~\eqref{eq:boundPair}. We subsequently estimate $T(n,D,q).$
The contribution $R(n,D,q)$ follows directly from \eqref{Red_complexity}.

\medskip
\noindent\textit{(i) The case $D$ grows sublinearly with $n$.}

\smallskip
\noindent
To estimate $N_1 \le \binom{n}{2}l_q(n,D-1)$, we show that
\[
l_q(n,D-1)\le\frac{D}{(n-D+1)}l_q(n,D).
\]
In fact, consider all admissible increments from $\gamma$ to $\gamma+e_i$. Since $|\gamma|=D-1$, at most $\lfloor\frac{D-1}{q-1}\rfloor \le D-1$ coordinates of $\gamma$ can be equal to $q-1$. Thus every $\gamma$ admits at least $n-D+1$ increments. On the other hand, every tuple of degree $D$ might come from at most $D$ tuples, one for each positive coordinate. Therefore
$(n-D+1)l_q(n,D-1)\le D\,l_q(n,D).$
Since $D=o(n)$ and $N= (1+ O(D/n ) )l_q(n,D)$, we have $N_1=O(nDN).$

For $N_2\leq\frac{1}{2} \sum_{t=1}^{D}[z^t]B_q(z)^n$, recall that
\[
[z^t]B_q(z)^n=
\sum_{\substack{|\delta|=t\\0\leq\delta_i<q}}
\prod_{i=1}^n(2\delta_i+1).
\]
Since $2a+1\le 3^a$ for every non-negative integer $a$, we have $\prod_{i=1}^n(2\delta_i+1)\le 3^{|\delta|}.$
Consequently,
\[
N_2 \le\frac12\sum_{t=1}^D3^t l_q(n,t) \le \frac{3^D}{2}N.
\]
The contribution $\mathbf{1}_{\{q\le D\}}nN$ is absorbed in $O(nDN).$
Overall this yields
\begin{equation}
\label{eq:rho-sublinear}
\rho =
O\left(
\frac{nD+3^D}{N}
\right).
\end{equation}
We have
\[
nD+3^D
\le D^2N^{1/D}+3^D
\le 2\cdot3^D N^{1/D},
\]
since $N\ge(n/D)^D$  and $D^2\le 3^D$. Also, $3^D=N^{o(1)}$ as
\[
\frac{\log(3^D)}{\log N}
\le \frac{\log 3}{\log(n/D)}
\longrightarrow 0.
\]

Substituting this in \eqref{eq:rho-sublinear}, it follows that the contribution $\rho N^4= O(N^{3+o(1)})$ when $D \to \infty$. 

\smallskip
\noindent
Now we estimate $T(n,D,q)=O( N^2 L_q(2n,D)).$ As $L_q(2n,D)\leq\binom{2n+D}{D}$ and $N\geq l_q(n,D)\geq\binom nD$, we have
\[
\frac{L_q(2n,D)}{N}\leq
\frac{\binom{2n+D}{D}}{\binom{n}{D}}=
\prod_{k=0}^{D-1}
2\left(1+\frac{D+k}{2(n-k)}\right).
\]
Using $1+x\leq e^x$ and $n-k\geq n-D+1$, we obtain
\[
\frac{L_q(2n,D)}{N}\leq
2^D \exp\left(
\frac{1}{2(n-D+1)}
\sum_{k=0}^{D-1}(D+k)
\right)
\leq
2^D\exp\left(
\frac{D^2}{n-D+1}
\right).
\]
Hence
\[
T(n,D,q) =
O\left(
2^D \exp\left(
\frac{D^2}{n-D+1}
\right) N^3 \right)=O(3^D N^3),
\]
since $\exp\left(\frac{D^2}{n-D+1}\right) \le (3/2)^D$ for $D=o(n)$.

\smallskip
\noindent
It holds that $T(n,D,q)+\rho N^4=O(N^{3+o(1)})$. Adding the contribution $R(n,D,q)= O(\mathbf 1_{\{q>D\}}qD\,N^2)$, this proves the first bound in (i).

\medskip
\noindent\textit{(ii) The case $D$ is bounded.}

\smallskip
\noindent
This case follows from the estimates in case (i). Here $l_q(n,d)= O(n^d)$  for every $d$, and therefore $N= O(n^D).$ Substituting this in \eqref{eq:rho-sublinear}  proves assertion (ii).
 
\medskip
\noindent\textit{(iii) The case $D=\Omega(n)$.}

\smallskip
\noindent
Clearly $N_1 \le \binom{n}{2}l_q(n,D-1)\le n^2N$.
Moreover, since $|\delta(\alpha,\beta)|\le D$ implies $|\alpha|,|\beta|\le D$, the coefficient interpretation of $B_q$ gives
\[
N_2\le \frac 12 \sum_{t=1}^D[z^t]B_q(z)^n\le \frac{N^2}{2}.
\]
Together with the contribution $\mathbf{1}_{\{q\le D\}}nN$, 
\[
\rho\le \frac{n^2N+N^2/2+nN}{N^2}\le\frac12+O\left(\frac{n^2}{N}\right) \le \frac 12 +o(1),
\]
as $N$ grows exponentially with $n$.
Really, since there exists a constant $c>0$ such that $D\ge cn$ for sufficiently large $n$,
choosing $0<c_<\min\{c,1\}$, we have 
\[N\ge\binom{n}{\lfloor c_0n\rfloor} \ge \left( \frac 1c_0 \right)^{\lfloor c_0n\rfloor}.\]

\smallskip
\noindent
Furthermore, $T(n,D,q)=O(N^2L_q(2n,D))=O(N^4)$, as $L_q(2n,D)\le L_q(n,D)^2$.

\smallskip
\noindent
Combining these estimates with $R(n,D,q)=O\left(\mathbf{1}_{\{q>D\}}qnN^2\right)$, proves (iii).

\end{proof}

\subsection{Memory complexity}\label{sec:mem}

We now estimate the memory requirements of the algorithm described in Section~\ref{sec:alg}, in terms of the number of $\mathbb{F}_q$ elements stored. We consider the storage needed for the Macaulay triangulations and the reductions of the field polynomials in the first stage, as discussed in Section~\ref{sec:newCompl}, and for the Gr\"{o}bner basis and the pairs processed in the second stage.

The Macaulay matrices can be triangulated incrementally, processing one row at a time. Each new row is reduced with the previous rows and stored only if it is linearly independent of them. Since there are $l_q(n,d)$ columns and, thus, at most $l_q(n,d)$ pivots,  the storage required at degree $d$ is $O(l_q(n,d)^2)$.

At degree $D$, we select $l_q(n,D)$ linearly independent rows of the homogeneous Macaulay matrix, forming an invertible matrix $C$. Considering the same rows with their lower degree terms gives a matrix $[C\mid E]$.
Multiplication by $C^{-1}$ yields $[I_{l_q(n,D)}\mid C^{-1}E],$ whose rows represent the polynomials $f_\rho=x^\rho+g_\rho$, with $\deg g_\rho<D$.
This part of the reconstruction requires $O(l_q(n,D)N)$ memory.
The reconstruction also accounts for the polynomials $f_\sigma$, whose leading monomials have degree $<D$. Since there are at most $L_q(n,D-1)$ such polynomials, each involving at most $N$ monomials, storing them requires $O(L_q(n,D-1)N)$ additional space.
The overall reconstruction of system \eqref{eq:s2} requires $O(N^2)$ memory.

Thus, processing the degrees incrementally, the memory required for the Macaulay triangulations and this reconstruction is
\[
O\left(
\max_{d\le D} l_q(n,d)^2+N^2
\right)
=O(N^2).
\]

When $q>D$, the reductions of the field polynomials can be performed one at a time. Since each intermediate polynomial contains at most $qL_q(n-1,D-1)$ monomial, these reductions require at most
\[
O\left(qL_q(n-1,D-1)\right)
\]
additional storage.

In the second stage, the current Gr\"{o}bner basis contains at most $N$ polynomials, each with at most $N$ monomials, and hence requires $O(N^2)$ storage. At most $N_1+N_2+\mathbf{1}_{\{q\le D\}}nN$ pairs need to be considered. Representing them by pairs of indices requires at most $O(N^2)$ additional storage.

Consequently, the overall memory complexity is bounded by
\[
O\left(
N^2+\mathbf{1}_{\{q>D\}}qL_q(n-1,D-1)
\right).
\]
For fixed $q$, the second contribution is absorbed into the former, and therefore the memory complexity is $O(N^2)$.

\bibliographystyle{unsrt}
\bibliography{bibl}

\begin{thebibliography}{10}

\bibitem{B65}
Bruno Buchberger.
\newblock {\em Ein Algorithmus zum Auffinden der Basiselemente des
  Restklassenringes nach einem nulldimensionalen Polynomideal}.
\newblock PhD thesis, Universit{\"a}t Innsbruck, 1965.

\bibitem{L83}
Daniel Lazard.
\newblock Gr{\"o}bner bases, gaussian elimination and resolution of systems of
  algebraic equations.
\newblock In J.~A. van Hulzen, editor, {\em Computer Algebra: EUROCAL 1983},
  volume 162 of {\em Lecture Notes in Computer Science}, pages 146--156.
  Springer, 1983.

\bibitem{F4}
Jean-Charles Faug{\`e}re.
\newblock A new efficient algorithm for computing {Gr{\"o}bner} bases ({F4}).
\newblock {\em Journal of Pure and Applied Algebra}, 139(1--3):61--88, 1999.

\bibitem{F5}
Jean-Charles Faug{\`e}re.
\newblock A new efficient algorithm for computing {Gr{\"o}bner} bases without
  reduction to zero ({F5}).
\newblock In {\em Proceedings of the 2002 International Symposium on Symbolic
  and Algebraic Computation}, pages 75--83. ACM, 2002.

\bibitem{XL}
Nicolas Courtois, Alexander Klimov, Jacques Patarin, and Adi Shamir.
\newblock Efficient algorithms for solving overdefined systems of multivariate
  polynomial equations.
\newblock In {\em Advances in Cryptology -- EUROCRYPT 2000}, volume 1807 of
  {\em Lecture Notes in Computer Science}, pages 392--407. Springer, 2000.

\bibitem{BFS03}
Magali Bardet, Jean-Charles Faug{\`e}re, and Bruno Salvy.
\newblock Complexity of {Gr{\"o}bner} basis computation of semi-regular
  overdetermined sequences over {$\mathbb{F}_2$} with solutions in
  {$\mathbb{F}_2$}.
\newblock Research Report 5049, INRIA, 2003.

\bibitem{BFS04}
Magali Bardet, Jean-Charles Faug{\`e}re, and Bruno Salvy.
\newblock On the complexity of {Gr{\"o}bner} basis computation of semi-regular
  overdetermined algebraic equations.
\newblock In {\em International Conference on Polynomial System Solving}, pages
  71--74, 2004.

\bibitem{XLdeg}
Jenny Yuan-Chun Yeh, Chen-Mou Cheng, and Bo-Yin Yang.
\newblock Operating degrees for {XL} vs. {$F_4/F_5$} for generic
  {$\mathcal{MQ}$} with number of equations linear in that of variables.
\newblock In Marc Fischlin and Stefan Katzenbeisser, editors, {\em Number
  Theory and Cryptography: Papers in Honor of Johannes Buchmann on the Occasion
  of His 60th Birthday}, volume 8260 of {\em Lecture Notes in Computer
  Science}, pages 19--33. Springer, Berlin, Heidelberg, 2013.

\bibitem{CG21}
Alessio Caminata and Elisa Gorla.
\newblock Solving multivariate polynomial systems and an invariant from
  commutative algebra.
\newblock In {\em Arithmetic of Finite Fields: WAIFI 2020}, volume 12542 of
  {\em Lecture Notes in Computer Science}, pages 3--36. Springer, 2021.

\bibitem{CG23}
Alessio Caminata and Elisa Gorla.
\newblock Solving degree, last fall degree, and related invariants.
\newblock {\em Journal of Symbolic Computation}, 114:322--335, 2023.

\bibitem{ST}
Igor Semaev and Andrea Tenti.
\newblock Probabilistic analysis on {Macaulay} matrices over finite fields and
  complexity of constructing {Gr\"o}bner bases.
\newblock {\em Journal of Algebra}, 565:651--674, 2021.

\bibitem{S25}
Flavio Salizzoni.
\newblock An upper bound for the solving degree in terms of the degree of
  regularity.
\newblock {\em Transactions on Mathematical Cryptology}, 5(1):1--7, 2025.

\bibitem{F4Tracer}
Robin Kouba, Vincent Neiger, and Mohab Safey El~Din.
\newblock A complexity analysis of the {F4} {Gr{\"o}bner} basis algorithm with
  tracer data, 2026.

\bibitem{GM}
R\"udiger Gebauer and H.~Michael M\"oller.
\newblock On an installation of {Buchberger}'s algorithm.
\newblock {\em Journal of Symbolic Computation}, 6(2):275--286, 1988.

\bibitem{B79}
Bruno Buchberger.
\newblock A criterion for detecting unnecessary reductions in the construction
  of {Gr\"o}bner bases.
\newblock In E.~W. Ng, editor, {\em Symbolic and Algebraic Computation}, pages
  3--21. Springer, 1979.

\bibitem{B85}
Bruno Buchberger.
\newblock Gr{\"o}bner bases: An algorithmic method in polynomial ideal theory.
\newblock In N.~K. Bose, editor, {\em Multidimensional Systems Theory}, pages
  184--232. D. Reidel Publishing Company, Dordrecht, 1985.

\end{thebibliography}

\appendix
\section{Exact counting formulas for processed pairs}
\label{app:exactPairs}

In this appendix we record the more precise counting formulas underlying Lemma~\ref{lem:N1N2}, together with equivalent generating function representations, and a more precise upper bound on the number of pairs involving field polynomials. For $N_1$, the formula below gives the exact number of pairs. For $N_2$, it gives the worst-case number of candidate pairs, obtained by assuming that every admissible monomial occurring in the count is the leading monomial of a polynomial available to the algorithm; the actual value of $N_2$ may therefore be smaller.

\subsection{Pairs contributing to $N_1$}

Recall that, for a pair contributing to $N_1$, the condition $|\alpha|=|\beta|=D$ and $|\delta|=D+1,$ gives $|\gamma|=D-1.$
Therefore, there exist two distinct indices $i,j$ such that $\alpha_i=\gamma_i+1,\ \beta_j=\gamma_j+1,$ with $0\le \gamma_i,\gamma_j\le q-2$.
If $\gamma_i+\gamma_j=k$, then the number of possibilities for  such $\gamma_i,\gamma_j$ is $l_{q-1}(2,k)$
whereas the remaining $n-2$ entries of $\gamma$ have total degree $D-1-k$ and can be chosen in $l_q(n-2,D-1-k)$ ways. Since there are $\binom{n}{2}$ possibilities for $i,j$, we obtain the exact number of pairs
\[
\binom{n}{2}
\sum_{k=
\max\{0,D-1-(n-2)(q-1)\}}^{
\min\{D-1,2(q-2)\}}
l_{q-1}(2,k)\,
l_q(n-2,D-1-k).
\]
The lower and upper limits are precisely the conditions ensuring that both factors in the summand are nonzero. That is $0 \le k \le 2(q-2) $ and $0 \le D-1-k \le (n-2)(q-1)$.

Equivalently, using coefficient extraction, the same quantity can be written as
\[
\binom n2
[z^{D-1}]
\left(\sum_{a=0}^{q-2}z^a\right)^2
\left(\sum_{a=0}^{q-1}z^a\right)^{n-2}.
\]

\subsection{Pairs contributing to $N_2$}

We showed that the number of ordered pairs $(\alpha,\beta)$ corresponding to $\delta$ is $\prod_{i=1}^{n}(2\delta_i+1).$
Let $s=|\operatorname{supp}(\delta)|.$
Among these ordered pairs, exactly $2^s$ have disjoint supports and have to be removed.
Indeed, for every $i\in\operatorname{supp}(\delta)$, one must choose $(\alpha_i,\beta_i)$ to be either  $ (\delta_i,0)$ or $(0,\delta_i).$
Remove also the pair $\alpha=\beta=\delta.$
It follows that the number of unordered, distinct pairs with non-disjoint supports and with maximum equal to $\delta$ is
\[
\frac{1}{2}
\left(
\prod_{i=1}^{n}(2\delta_i+1)-2^s-1
\right).
\]

Notice also that the condition that at least one of the two polynomials has degree strictly smaller than $D$ is automatic in the above count.

Summing over all nonzero exponent tuples $\delta$ of degree at most $D$ gives
\[
\frac{1}{2}
\sum_{\substack{
0<|\delta|\le D\\
0\le \delta_i<q
}}
\left(
\prod_{i=1}^{n}
(2\delta_i+1)-2^{|\operatorname{supp}(\delta)|}-1
\right).
\]

For the equivalent generating function representation, define
\[
A_q(z)=\sum_{j=0}^{q-1}z^j,
\qquad
B_q(z)=\sum_{j=0}^{q-1}(2j+1)z^j,
\qquad
C_q(z)=1+2\sum_{j=1}^{q-1}z^j.
\]
Then $B_q(z)^n$ counts all ordered pairs according to the degree of their coordinatewise maximum, $C_q(z)^n$ counts those with disjoint supports, and $A_q(z)^n$ counts the diagonal pairs. Hence the preceding quantity is equivalently
\[
\frac12\sum_{t=1}^{D}
[z^t]\left(B_q(z)^n-C_q(z)^n-A_q(z)^n\right).
\]

\subsection{Pairs involving field polynomials}

It remains to count the pairs involving a field polynomial. Such pairs need to be considered only when $q\le D$.
Pairs involving two distinct field polynomials have coprime leading monomials and can be discarded by Buchberger's criterion (C1).

Consider a pair formed by a field polynomial with leading monomial $x_j^q$ and a polynomial whose leading monomial is $x^\beta$. For the two supports to intersect, one must have $1\le\beta_j\le q-1$. Moreover,
\[
|\delta|=q+\sum_{i\ne j}\beta_i.
\]
By Lemma~\ref{lem:redundantField}, such a pair needs to be considered only if either $|\delta|\le D$, or $|\beta|=D$ and $|\delta|= D+1$.

In the first case, the condition $|\delta|\le D$ is equivalent to
\[
\sum_{i\ne j}\beta_i\le D-q.
\]
For a fixed $j$, there are $q-1$ choices for $\beta_j$, and the remaining coordinates can be chosen in at most $L_q(n-1,D-q)$ ways. Summing over the $n$ field polynomials gives
\[
n(q-1)L_q(n-1,D-q).
\]

In the second case, since $|\beta|=D$ and and $|\delta|= D+1$ imply $\beta_j=q-1$, we have
\[
\sum_{i\ne j}\beta_i=D-q+1.
\]
For fixed $j$, there are at most $l_q(n-1,D-q+1)$ possible $\beta$. Thus, this contributes exactly
\[
n\,l_q(n-1,D-q+1)
\]
pairs.

Therefore, the number of selected pairs involving a field polynomial is at most
\[
\mathbf{1}_{\{q\le D\}}\,n
\left(
(q-1)L_q(n-1,D-q)
+l_q(n-1,D-q+1)
\right).
\]

\section{An exponential improvement for large degrees of regularity}
\label{app:largeD}
\noindent\textit{Statement on AI use.}
We used ChatGPT (OpenAI, GPT-5.6 Sol) as a tool to explore alternative proof strategies for the result presented in this section.
The final mathematical arguments were developed and verified by the authors, who take full responsibility for their correctness.

\medskip
Recall that the total complexity is bounded by $O(T(n,D,q)+R(n,D,q)+\rho N^4)$, where $N=L_q(n,D),\ \rho = (N_1+N_2+\mathbf{1}_{\{q\le D\}}\,nN)/N^2$ and
\[
T(n,D,q)=
O\left(N^2L_q(2n,D)\right),
\quad
R(n,D,q)=
O\left(\mathbf 1_{\{q>D\}}\frac{nqD}{n+D}N^2\right).
\]
We refine the complexity bound for $D=\Omega(n)$ in the range $\frac12\le\lambda\le\theta_q-\varepsilon,$ where
\[
\lambda=\frac{D}{n(q-1)},\qquad
\theta_q=\frac{4q+1}{6q},
\]
and $\varepsilon>0$ is fixed.

\begin{thm}\label{thm:complLargeD}
In the above range, the total complexity of the algorithm is $O(c^nN^4)$ for some constant $c<1$, independent of $n$ and $q$.
\end{thm}

\begin{proof}
The components map
\[
(\alpha_1,\ldots,\alpha_n)\mapsto
(q-1-\alpha_1,\ldots,q-1-\alpha_n)
\]
pairs exponent tuples whose degrees sum to $n(q-1)$.
Hence at least half of all $q^n$ tuples have degree at most $n(q-1)/2$. Hence $N\ge q^n/2$, since $\lambda\ge1/2$.

Let $\alpha$ and $\beta$ be independent uniformly distributed exponent tuples in $\{0,\ldots,q-1\}^n$, and set $X_i=\max(\alpha_i,\beta_i)/(q-1)$. The variables $X_i$ are independent, belong to $[0,1]$, and satisfy $\mathbb E[X_i]=\theta_q$. Indeed, For a fixed coordinate $i$, exactly $2j+1$ ordered pairs $(\alpha_i,\beta_i)$ satisfy $\max(\alpha_i,\beta_i)=j,\ 0\le j<q$. Thus $\Pr(\max(\alpha_i,\beta_i)=j)=(2j+1)/q^2,$ and so
\[
\mathbb E[X_i]=
\frac1{q^2(q-1)}
\sum_{j=0}^{q-1}j(2j+1)
=\theta_q.
\]
By Hoeffding's inequality,
\[
\Pr\left(\sum_{i=1}^n X_i\le n\lambda\right)
\le \exp\left(-2n(\theta_q-\lambda)^2\right)
\le e^{-2\varepsilon^2n}.
\]
Recall that $B_q(z)=\sum_{j=0}^{q-1}(2j+1)z^j$ and, for each $t$, the coefficient $[z^t]B_q(z)^n$ counts the ordered pairs of exponent tuples $(\alpha,\beta)$ such that $\sum_{i=1}^n\max(\alpha_i,\beta_i)=t.$
Since $\sum_{i=1}^n\max(\alpha_i,\beta_i)=(q-1)\sum_{i=1}^n X_i$ and $D=n(q-1)\lambda,$ we have
\[
\sum_{t=0}^D[z^t]B_q(z)^n=q^{2n}\Pr\left(\sum_{i=1}^n X_i\le n\lambda\right).
\]
Consequently,
\[
\frac12\sum_{t=1}^D[z^t]B_q(z)^n
\le \frac12q^{2n}e^{-2\varepsilon^2n}.
\]
Using \eqref{eq:boundPair} and $N\ge q^n/2$, we obtain
\[
\rho=\frac{N_1+N_2}{N^2}
\le 2e^{-2\varepsilon^2n}
+O\left(\frac{n^2}{N}\right)
=O(e^{-2\varepsilon^2n}).
\]

To estimate the contribution $T(n,D,q)$, let $Y_i=(\alpha_i+\beta_i)/(q-1)$. These variables are independent, belong to $[0,2]$, and have mean $1$. Since $\lambda\le\theta_q\le3/4$, Hoeffding's inequality gives
\[
\begin{aligned}
L_q(2n,D)
&=q^{2n}\Pr\left(
\sum_{i=1}^nY_i\le n\lambda
\right)\\
&\le q^{2n}
\exp\left(-\frac{n}{2}(1-\lambda)^2\right)\\
&\le q^{2n}e^{-n/32}
\le 4N^2e^{-n/32}.
\end{aligned}
\]
Therefore, $T(n,D,q)=O(e^{-n/32}N^4).$

Finally, $\lambda\ge1/2$ implies $q\le D$ for sufficiently large $n$, so the additional cost $R(n,D,q)$ to reduce field polynomials vanishes. Combining the two stages gives $O(c^nN^4)$ for some $c<1$.
\end{proof}

\end{document}